%% file: main.tex
\documentclass[10pt, conference]{IEEEtran}

\input{Preamble}

\input{Commands}

\glsenableentrycount
\input{Acronyms}

\IEEEoverridecommandlockouts\IEEEpubid{\makebox[\columnwidth]{ 979-8-3315-2037-3/25/\$31.00 ©2025 IEEE \hfill} \hspace{\columnsep}\makebox[\columnwidth]{ }}
\begin{document}
\bstctlcite{CustomBSTcontrol}

\sisetup{
detect-weight,
mode=text,
range-units=single, 
range-exponents=combine-bracket,
range-phrase={\,--\,}
}

\title{A Multi-Modal IoT Node for Energy-Efficient Environmental Monitoring with Edge AI Processing
\thanks{We thank Philipp Schilk for his valuable contributions to the research project. This work was supported by the ETH-Domain
Joint Initiative program (Project UrbanTwin).}}

\author{
\IEEEauthorblockN{
        Philip Wiese\IEEEauthorrefmark{1}~\orcidlink{0009-0001-7214-2150} ,
        Victor Kartsch\IEEEauthorrefmark{1}~\orcidlink{0000-0003-3325-6347},
        Marco Guermandi\IEEEauthorrefmark{2}~\orcidlink{0000-0001-9801-5769},
        Luca Benini\IEEEauthorrefmark{1}\IEEEauthorrefmark{2}~\orcidlink{0000-0001-8068-3806}
}
\IEEEauthorblockA{
       \IEEEauthorrefmark{1}\textit{Integrated Systems Laboratory (IIS), ETH Zurich}, Switzerland \\
    \IEEEauthorrefmark{2}\textit{Department of Electrical, Electronic and Information Engineering (DEI), University of Bologna}, Italy 
}
\IEEEauthorblockA{
     \{wiesep, victor.kartsch, lbenini\}@iis.ee.ethz.ch, marco.guermandi@unibo.it
  }
}

\maketitle

\IEEEpubid{\begin{minipage}{\textwidth}
\centering \footnotesize
\vspace{7em} 
\copyright 2025 IEEE. Personal use of this material is permitted. Permission from IEEE must be obtained for all other uses, in any current or future media, including reprinting/republishing this material for advertising or promotional purposes, creating new collective works, for resale or redistribution to servers or lists, or reuse of any copyrighted component of this work in other works.
\end{minipage}}

\IEEEpubidadjcol


\input{src/00_Abstract}

\IEEEpeerreviewmaketitle


\input{src/10_Introduction}
\input{src/20_RelatedWork}

\input{src/30_Architecture}
\input{src/40_Occupancy}

\input{src/50_Results}
\input{src/70_Discussion}
\input{src/80_Conclusion}
\input{src/90_Acknowledgement}

\bibliographystyle{IEEEtran}
\bibliography{\jobname, ./references}

\vfill
\cleardoublepage


\end{document}

%% file: Preamble.tex
\usepackage{graphicx}
\usepackage[hidelinks]{hyperref}
\usepackage[
free-standing-units,
per-mode=symbol,
mode=text
]{siunitx}
\usepackage{amssymb,amsfonts}
\usepackage{amsmath}
\usepackage[export]{adjustbox}
\usepackage{float}
\usepackage{textgreek}
\usepackage{url}
\usepackage[noadjust]{cite}
\usepackage{orcidlink}
\usepackage[acronym]{glossaries}
\usepackage{tabularx}
\usepackage{enumerate}
\usepackage{threeparttable}
\usepackage{multirow}
\usepackage{booktabs} 
\usepackage{xcolor}   
\usepackage{pifont}
\usepackage{verbatim}
\usepackage{cleveref}
\usepackage{soul}
\usepackage[
caption=false,
margin=5pt]{subfig}
\usepackage[version=4]{mhchem}  


%% file: Commands.tex
\DeclareSIUnit\gateequivalent{GE}
\DeclareSIUnit\operation{OP}
\DeclareSIUnit\decibelmeter{dBm}
\DeclareSIUnit\fps{FPS}
\DeclareSIUnit{\amperehour}{Ah}

\soulregister\gls7
\soulregister\cgls7
\soulregister\cglspl7
\soulregister\Gls7
\soulregister\cGls7
\soulregister\cGlspl7
\soulregister\acrshort7
\soulregister\acrlong7
\soulregister\Acrshort7
\soulregister\Acrlong7
\soulregister\Acrlong7
\soulregister\SI7
\soulregister\num7
\soulregister\cref7
\soulregister\Cref7
\soulregister\cite7
\soulregister\textcolor7

\newcommand{\todo}[1]{}
\newcommand{\R}[1]{#1}

\newcommand{\revA}[1]{#1}
\newcommand{\revB}[1]{#1}

%% file: Acronyms.tex
\newacronym{EM}{EM}{environmental monitoring}
\newacronym{ml}{ML}{machine learning}
\newacronym{svm}{SVM}{support vector machine}
\newacronym{rf}{RF}{random forest}
\newacronym{dcv}{DCV}{demand-controlled ventilation}
\newacronym{cnn}{CNN}{convolutional neural network}
\newacronym{iaq}{IAQ}{indoor air quality}
\newacronym{voc}{VOC}{volatile organic compound}
\newacronym{soa}{SoA}{state-of-the-art}
\newacronym{pulp}{PULP}{parallel ultra-low power}
\newacronym{soc}{SoC}{system-on-chip}
\newacronym{IoT}{IoT}{Internet of Things}
\newacronym{NB-IoT}{NB-IoT}{Narrowband IoT}
\newacronym{LoRa}{LoRa}{Long Range Communication}
\newacronym{lstm}{LSTM}{long short-term memory}
\newacronym{xgbrf}{XGBRF}{random forest with XGBoost}
\newacronym{pmic}{PMIC}{power management integrated circuit}
\newacronym{ai}{AI}{artificial intelligence}
\newacronym{nms}{NMS}{non-maximum suppression}
\newacronym{ann}{ANN}{artificial neural networks}
\newacronym{genalgo}{GA}{genetic algorithms}
\newacronym{BLE}{BLE}{Bluetooth Low Energy}
\newacronym{uart}{UART}{universal asynchronous receiver-transmittery}
\newacronym{IoU}{IoU}{intersection over union}
\newacronym{mAP}{mAP}{mean average precision}
\newacronym{mAP50}{mAP\textsubscript{50}}{\cgls{mAP} at an \cgls{IoU} threshold of 0.50}
\newacronym{mAP5095}{mAP\textsubscript{50-95}}{average of the \cgls{mAP} at a varying \cgls{IoU} threshold ranging from 0.50 to 0.95}
\newacronym{PIR}{PIR}{passive infrared}
\newacronym{IMU}{IMU}{inertial measurement unit}
\newacronym{ulp}{ULP}{ultra-low-power}
\newacronym{mcu}{MCU}{microcontroller}
\newacronym{wsn}{WSN}{wireless sensor network}
\newacronym{dsp}{DSP}{digital signal processor}
\newacronym{fpga}{FPGA}{field-programmable gate array}
\newacronym{PTQ}{PTQ}{post-training quantization}

%% file: src/00_Abstract.tex
\begin{abstract}
The widespread adoption of \gls{IoT} technologies has significantly advanced \gls{EM} by enabling cost-effective and scalable sensing solutions. 
\revB{Concurrently, \gls{ml} and \gls{ai} are introducing powerful tools for analyzing complex environmental data efficiently. 
However, current IoT \gls{EM} platforms are typically limited to a narrow set of sensors and lack the computational capabilities to support advanced \gls{ml} and \gls{ai} on the edge.
To overcome these limitations, we introduce a compact (17x38\,mm\textsuperscript{2}), multi-modal, microcontroller based environmental IoT node integrating 11 sensors, including \ce{CO2} concentration, \glspl{voc}, light intensity, UV radiation, pressure, temperature, humidity, RGB camera, and precise geolocation through a GNSS module.
It features GAP9, a parallel ultra-low-power system-on-chip, enabling real-time, energy-efficient on-device \gls{ml} processing.}
We implemented a YOLOv5-based occupancy detection pipeline (\R{\SI{0.3}{\mega\relax}} parameters, \R{\SI{42}{MOP}} per inference), demonstrating \R{\SI{42}{\percent}} energy savings over raw data streaming.
\revB{Additionally, we present a smart \gls{iaq} monitoring setup that combines occupancy detection with adaptive sample rates, achieving operational times of up to \R{\SI{143}{\hour}} on a single compact \R{\SI{600}{\milli\amperehour}}, \R{\SI{3.7}{\volt}} battery.}

\end{abstract}
\renewcommand\IEEEkeywordsname{Keywords}

\glsresetall

%% file: src/10_Introduction.tex
\section{Introduction}
\label{sec:intro}

The widespread adoption of \cgls{IoT} technologies has significantly enhanced \cgls{EM}, offering scalable and cost-effective solutions for data-driven decision-making in the face of climate change~\cite{UN_Climate_DegreesMatter}. 
Recent progress in sensing and embedded processing has enabled \cgls{IoT}-based monitoring across diverse applications, from tracking greenhouse gases guiding policy to assessing air and water quality for public health and, more recently, supporting urban sustainability initiatives through digital twin modeling and smart city planning \cite{urbantwin2023}.

Advances in sensor technology have expanded the capabilities of \cgls{IoT}-based \cgls{EM} systems by enabling low-cost (under \SI{20}{EUR}) and accurate sensor nodes. 
However, capturing comprehensive environmental information requires multiple sensing modalities, increasing system complexity and deployment costs \cite{tang2023comparative, moursi2021iot}.
This highlights the need for efficient, scalable, and flexible platforms capable of multi-modal sensing to address the growing demands of digitalized environmental monitoring.

\cGls{ml} offers new opportunities for enhancing \cgls{EM} by enabling scalable and accurate analysis of complex sensor data. 
Traditional models, such as \cglspl{svm} and \cglspl{rf}, alongside deep learning frameworks like \cglspl{cnn}, have proven effective in applications including land-use classification, compliance inspection targeting, and crowd density estimation. 

Moreover, \cgls{ml} contributes to system scalability and efficiency by reducing data transmission needs through on-device inference, which is especially critical for high-bandwidth sensors such as cameras and microphones~\cite{murshed2021machine, 10595987, busia2025endoscopy}.
A notable example is \cgls{iaq} monitoring, where \cgls{ml} can infer occupancy with cameras or detect pollutants like \cglspl{voc}, dust, and \ce{CO2}, which are linked to health symptoms, including fatigue, headaches, and respiratory issues. 
In this context, \cgls{ml} reduces the volume of transmitted data and addresses privacy concerns by keeping sensitive processing local~\cite{9893137, sabovic2023towards}.

Combining \cgls{ml} with hardware platforms capable of efficient multi-modal sensing opens the door to more autonomous and intelligent monitoring systems. 
These systems can go beyond passive observation to predict environmental trends and enable proactive interventions. 
Realizing this vision requires addressing several open challenges: integrating multiple sensing modalities into compact, low-power platforms, reducing the memory and computational complexity of current \cgls{ai} algorithms, and finally, increasing the processing capabilities of resource-constrained \cgls{IoT} devices.
In fact, most platforms surveyed~\cite{karray2018comprehensive, wu2022microcontroller} lack efficient computational capabilities for edge-AI processing and/or do not include a multi-modal sensing approach.

To fill this gap, we present a flexible \cgls{mcu} based environmental monitoring system supporting flexible \cgls{ulp} onboard \cgls{ai}, capturing a wide range of environmental parameters, including \ce{CO2} concentration, light intensity, UV radiation, \cglspl{voc}, pressure, temperature, and humidity.
It also integrates an RGB camera for visual sensing and a GNSS module for precise geolocation. 
At its core is GAP9, an energy-efficient \cgls{pulp} \cgls{soc}, which enables local inference via edge \cgls{ai}. 
Our system bridges the gap between general-purpose \cgls{IoT} nodes and existing environmental sensing platforms by supporting diverse applications by integrating a highly programmable parallel \cgls{ulp} \cgls{soc}, offering enough computational power for \cgls{ai}-based processing at high energy efficiency.
It ensures enhanced privacy, reduces reliance on external infrastructure, lowers overall system costs, and provides increased robustness and safety.
We demonstrate its effectiveness through an on-device deployment of YOLOv5 for indoor occupancy detection and multi-modal \cgls{iaq} monitoring.

This approach lays the groundwork for predictive \cgls{iaq} monitoring, where embedded models can forecast key environmental metrics and trigger actions such as ventilation control or early-warning alerts.
This will offer the possibility of proactive control strategies to mitigate air pollution better, optimize ventilation, and trigger early-warning alarms. 
Specifically, our contributions are as follows:

\begin{itemize}
    \item A compact, modular, multi-modal \cgls{mcu}-based environmental monitoring platform integrating 11 diverse sensors, including gas, pressure, temperature, visual, and motion, to provide comprehensive environmental awareness and operate autonomously over an extended time period\footnote{SENSEI Sensor Shield: \url{https://github.com/pulp-bio/sensei-sensor-shield}}. 

    \item Integration of GAP9\footnote{Through a pre-existing board \cite{frey2025biogap}, not developed as part of this work}, a flexible, energy-efficient \acrlong{pulp} \cgls{soc}, enabling local and privacy-preserving execution of advanced \cgls{ml} algorithms through embedded edge \cgls{ai}.

    \item Extensive performance characterization of an optimized YOLOv5 model for occupancy detection deployed on GAP9, highlighting key trade-offs between energy consumption and inference accuracy, and demonstrating a \R{\SI{42}{\percent}} energy reduction compared to raw image streaming.

    \item Evaluation of an autonomous, full-stack \cgls{iaq} monitoring application, combining real-time occupancy detection with adaptive environmental sensing. Using the most accurate network configuration, we demonstrate continuous operation for up to \R{\SI{143}{\hour}} on a single compact \R{\SI{600}{\milli\amperehour}}, \SI{3.7}{\volt} battery.

\end{itemize}

%% file: src/20_RelatedWork.tex
\section{Related Work}
\label{sec:background}

\begin{figure*}[t]
    \centering
    \subfloat[
       SENSEI Base Board with a dual-SoC architecture for near-sensor processing. It combines the GAP9 for parallel computation and the nRF5340 for low-power wireless connectivity. The board includes a high-density board-to-board connector for stacking application-specific shields.
        \label{fig:base_board}
    ]{
        \includegraphics[width=0.47\textwidth]{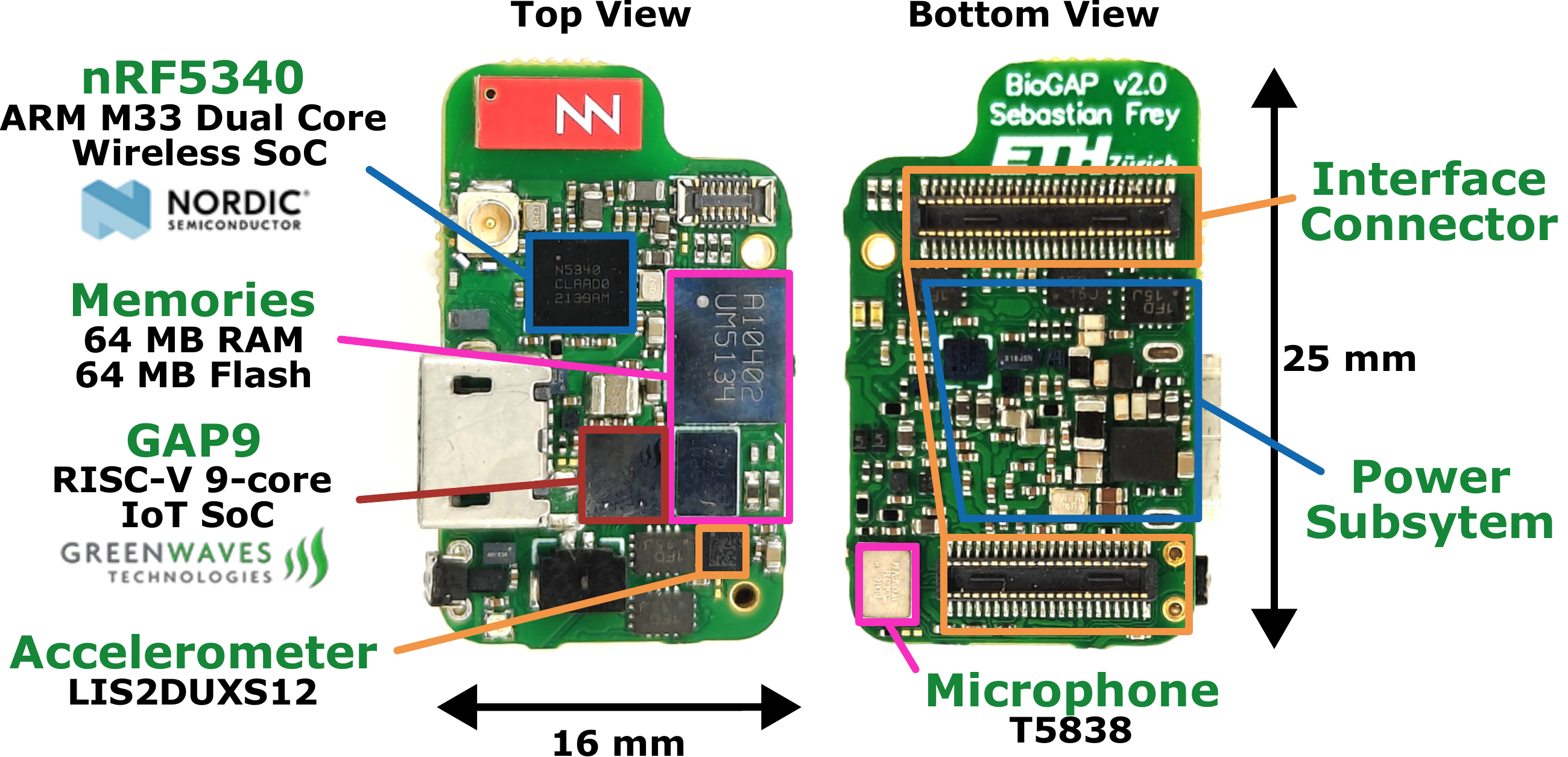}
    }
    \hspace{1em}
    \subfloat[
        SENSEI Environmental Sensing Shield featuring a wide range of sensors for air quality, motion, light, and location. Designed for long-term data acquisition and contextual monitoring, it interfaces seamlessly with the base board via board-to-board connectors.
        \label{fig:sensor_shield}
    ]{
        \includegraphics[width=0.47\textwidth]{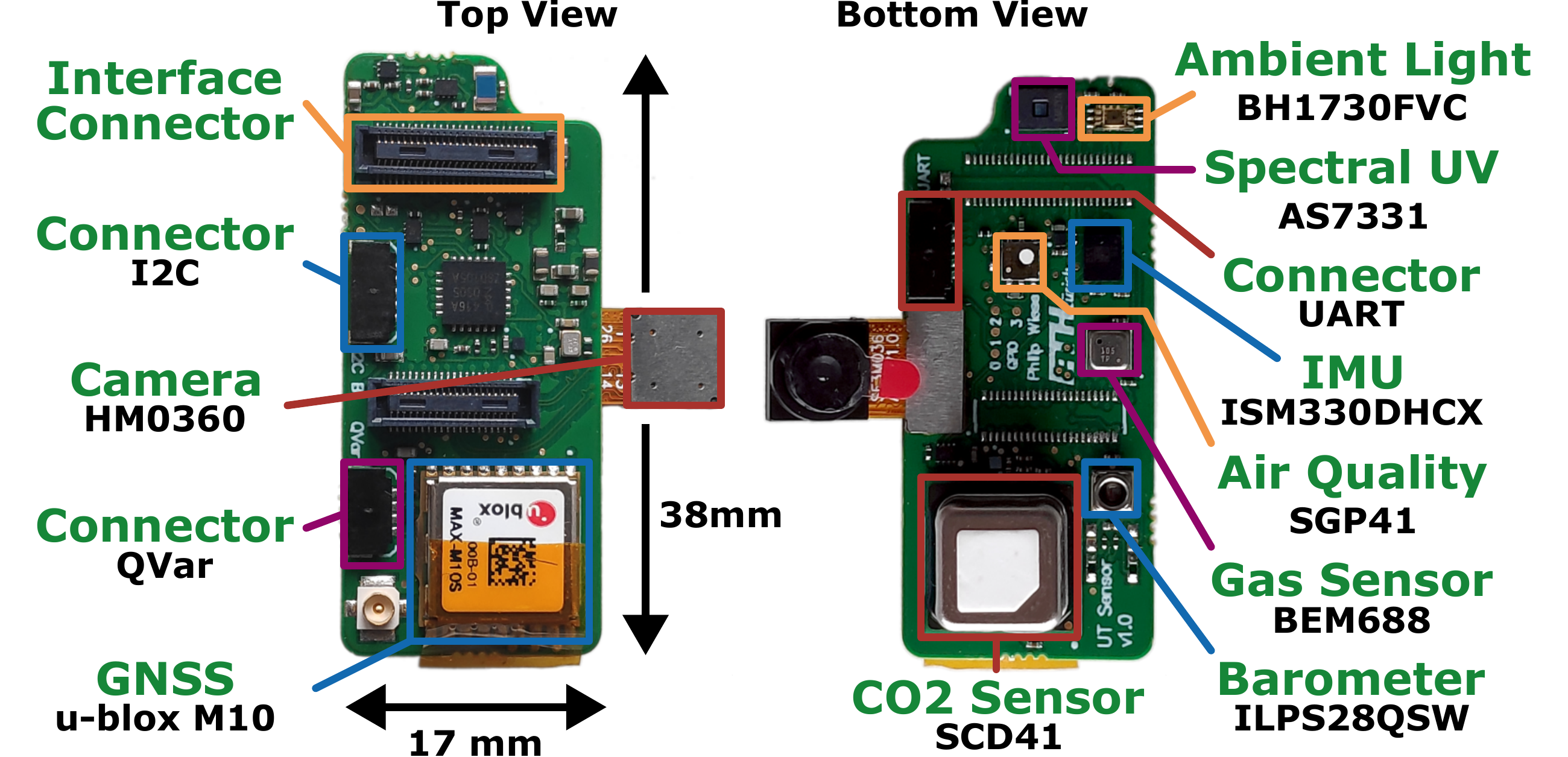}
    }
    \caption{Our platform combines a compact dual-SoC base board (a) with a modular sensing shield (b) to enable energy-efficient, near-sensor computation for environmental and contextual monitoring.}
    \label{fig:sensei_platform}
    \vspace{-1em}
\end{figure*}

\cGls{IoT} is transforming \cgls{EM} by integrating advanced sensing with energy-efficient embedded processing~\cite{quy2022iot}. 
These developments have modernized key sectors, including agriculture~\cite{chamara2022ag}, healthcare~\cite{frey2023biogap}, and infrastructure~\cite{9410570}, and have enabled large-scale tracking of environmental indicators related to climate change~\cite{anh2021wireless}. 
When combined with emerging communication protocols such as \acrfull{NB-IoT}, \acrfull{LoRa}, and \cgls{BLE}, \cgls{IoT} systems offer promising solutions for distributed and decentralized data processing~\cite{9938439}.

Several platforms have been proposed for applying \cgls{IoT} to \cgls{EM}.
Almalki et al.~\cite{almalki2021low} presented a system for farm monitoring that measures soil moisture, temperature, and humidity using drone-assisted wireless data collection.
Islam et al.~\cite{islam2023monitoring} developed a platform for real-time water quality monitoring in fish ponds, using sensors for pH, temperature, turbidity, conductivity, and dissolved oxygen, and applied \cgls{ml} techniques to predict fish survival.
Sindhwani et al.~\cite{Sindhwani2022Thingspeak} introduced an \cgls{IoT}-based radiation detection node using the ThingSpeak platform, which includes sensors for temperature, humidity, gas, smoke, sound, and pressure. 
Although these systems provide useful sensing capabilities, they lack on-device processing and thus require centralized computation, increasing latency, energy consumption, and data privacy risks.

Furthermore, besides specific examples, recent surveys have provided comprehensive overviews of platforms used in \cglspl{wsn} and visual sensing. 
Karray et al.~\cite{karray2018comprehensive} presented an extensive review of \cgls{wsn} hardware, including \cglspl{mcu}, \cglspl{dsp}, and \cglspl{fpga}. 
Khalifeh et al.~~\cite{s22228937} reviewed specifically \cgls{mcu}-based sensor nodes, while in \cite{costa2020visual} the author focuses on visual sensor platforms for \cgls{IoT} applications. 
However, while these reviews highlight their flexibility in sensing tasks, they also show that current platforms are unsuitable for energy-efficient extreme edge computing implementation of the new generation of \cgls{ai} algorithms \cite{murshed2021machine, song2024tinyml}. 
Critical gaps must be addressed to enable effective AI integration within the environmental sensing ecosystem. Key challenges include the need for on-device \cgls{ai} inference, programmable acceleration, multi-modal environmental and visual sensing support, and ultra-low-power operation suitable for battery-powered devices. 

In fact, the need for efficient edge processing comes along with new \cGls{ml} and \cgls{ai}, which have further advanced the field by enabling complex pattern recognition and forecasting capabilities. 
Garcia et al.~\cite{garcia2022smart} applied offline Gaussian Process Regression to predict industrial air quality based on pollutants such as \ce{SO2}, \ce{NO2}, \ce{O3}, and \ce{CO}.
Essa et al.~\cite{essa2022deep} explored the use of three \cgls{lstm} based models to forecast thunderstorm severity using lightning frequency from remote sensing data. 
Similarly, Monta{\~n}a et al.~\cite{Montana2024ThunderstormCNN} proposed a 1D \cgls{cnn} with attention mechanisms for thunderstorm days prediction, achieving an F1 score of \SI{79.6}{\percent} using weather and lightning features.
Al-Khafajiy et al.~\cite{Al-Khafajiy2022IAQ} used an \cgls{lstm} network to model \cgls{iaq} trends from sensors measuring temperature, humidity, pressure, \ce{CO2}, \ce{CO}, and particulate matter (PM2.5).
Sharma et al.~\cite{Sharma2022SmartAgri} introduced a hybrid framework combining \cgls{ann} and \cgls{genalgo} for wheat yield prediction from image data, achieving over \SI{97}{\percent} accuracy.
While these approaches demonstrate strong predictive performance, they generally overlook the computational and memory constraints of embedded platforms, limiting their applicability to battery-operated, stand-alone systems.

Some studies have started addressing this gap by incorporating edge processing. 
Moursi et al.~\cite{moursi2021iot} proposed a system in which sensor data is transmitted to a Raspberry Pi 4 for air quality prediction. 
Shadrin et al.~\cite{Shadrin2020EmbeddedAI} implemented an \cgls{lstm}-based model for monitoring leaf area dynamics in tomato plants using a Raspberry Pi 3B and an Intel Movidius Neural Compute Stick (NCS2). 
Albanese et al.~\cite{albanese2021automated} developed a \cgls{cnn}-based pest detection platform for precision agriculture using a Raspberry Pi 3 and NCS2. 
Yen et al.~\cite{yen2023adaptive} introduced a people-counting system based on a Raspberry Pi 4 and a fisheye camera, employing a modified YOLOv4-tiny model optimized for embedded deployment. 
However, these platforms are relatively power-hungry and unsuitable for long-term deployment in low-power scenarios due to their high energy consumption and limited integration of sensing and processing.

To address these limitations, we introduce a multi-modal sensing platform with embedded \cgls{ai} capabilities based on the ultra-low-power GAP9 \cgls{soc}. 
The system combines energy-efficient data acquisition with local inference, enabling both real-time occupancy monitoring via a deployed \cgls{cnn} and environmental data collection for future predictive modeling. 
This approach supports the development of autonomous, privacy-preserving, and battery-efficient \cgls{EM} systems suitable for long-term deployment.

%% file: src/30_Architecture.tex
\section{Hardware Architecture}
\label{sec:architecture}


SENSEI is a modular, ultra-low-power platform designed for edge \cgls{ai} and near-sensor signal processing. 
It combines real-time and energy-efficient computation with advanced connectivity and flexible sensor integration, enabling scalable, privacy-aware \cgls{EM} across diverse deployment scenarios.

\subsection{SENSEI Base Board}

The SENSEI Base Board\footnote{SENSEI Base Board: \url{https://github.com/pulp-bio/sensei-base-board}}, shown in \Cref{fig:base_board}, is an open-source hardware platform developed within the PULP-Platform \cite{frey2025biogap}. 
It serves as the computational and communication backbone of our system and was adopted in this work due to its support for energy-efficient edge processing and modular extensibility.
It integrates two key processing units and a flexible power management IC in a compact \SI[parse-numbers=false]{16\times25}{\square\milli\meter} form factor optimized for energy-efficient edge computation and modular expansion.

\begin{itemize}
    \item \textbf{GAP9 SoC:} A 9-core RISC-V \cgls{soc} operating at up to \SI{370}{\mega\hertz}, equipped with \SI{128}{\kilo\byte} of L1 and \SI{1.5}{\mega\byte} of L2 on-chip memory. It features the NE16 hardware accelerator for low-power deep learning inference. The \cgls{soc} integrates \SI{2}{\mega\byte} of non-volatile MRAM, and is extended with external \SI{64}{\mega\byte} RAM and \SI{64}{\mega\byte} Flash for memory-intensive workloads.

    \item \textbf{nRF5340 SoC:} A dual-core Arm Cortex-M33 system offering multi-protocol wireless connectivity, including \cgls{BLE}~5.4, IEEE 802.15.4, and proprietary 2.4\,GHz protocols. The application core includes \SI{1}{\mega\byte} Flash and \SI{512}{\kilo\byte} RAM, with an additional \SI{16}{\mega\byte} of external RAM for extended buffering and data handling.

    \item \textbf{MAX77654 PMIC:} A flexible power management IC that supports multiple programmable voltage rails, battery charging, and protection mechanisms against overcurrent and overheating, ensuring safe operation under various environmental conditions.

    \item \textbf{On-Board Sensors:} The board includes the \textit{LIS2DUXS12} ultra-low-power accelerometer for always-on activity monitoring and the \textit{T5838} PDM microphone for acoustic sensing.
\end{itemize}
A high-density board-to-board connector facilitates stacking with application-specific expansion modules.
The board also provides multiple peripheral interfaces and GPIO headers for flexible hardware integration, supporting long-term, autonomous deployments in various settings.

\subsection{SENSEI Environmental Sensing Shield}




As part of this work, we designed the SENSEI Environmental Sensing Shield, shown in \Cref{fig:sensor_shield}, to complement the base board by integrating a comprehensive set of sensors tailored for environmental and contextual data acquisition. 
The shield design enables seamless interoperability with the base board via board-to-board connectors and supports adaptive, context-aware applications through on-device processing.
It was developed to meet the specific requirements of multi-modal, edge-\cgls{ai}-enabled monitoring, balancing sensing versatility, energy efficiency, and modularity for long-term deployment scenarios. 
\begin{itemize}
    \item \textbf{Optical Sensors:} The \textit{AS7331} enables ultraviolet (UVA/B/C) sensing, while the \textit{BH1730FVC} captures ambient visible and infrared light for illumination-aware applications. The \textit{HM0360} camera provides \SI[parse-numbers=false]{640\times480}{} resolution at \SI{60}{\fps}, supporting computer vision tasks such as occupancy detection and environmental classification.
    
    \item \textbf{Environmental Sensors:} The \textit{BME680} measures temperature, humidity, and gas resistance. The \textit{SGP41} monitors volatile organic compounds (\cglspl{voc}), while the \textit{SCD41} provides precise \ce{CO2} measurements. The \textit{ILPS28QSW} barometric pressure sensor supports altitude estimation and weather tracking.

    \item \textbf{Motion and Positioning:} The \textit{ISM330DHCX} is a high-performance 6-DoF \acrfull{IMU} combining a 3-axis accelerometer and gyroscope for motion and vibration sensing. The \textit{MAX-M10S} GNSS module enables accurate geolocation and time synchronization of sensor data.

    \item \textbf{External Interfaces:} I\textsuperscript{2}C, UART, and GPIO connectors allow seamless integration of additional peripherals and application-specific sensors.

\end{itemize}

The shield is designed for high-resolution, multi-modal data acquisition in both indoor and outdoor settings and enables adaptive and localized intelligence directly at the edge.

%% file: src/40_Occupancy.tex
\section{Indoor Occupancy Detection}
\label{sec:occupancy_detection}



To demonstrate the edge \cgls{ai} capabilities of our platform, we implement and evaluate an embedded pipeline for indoor occupancy detection using RGB camera inputs. 
For this purpose, we adopt YOLOv5, a widely used object detection architecture based on a convolutional backbone and detection head optimized for real-time performance. 
To meet the resource constraints of our platform, we use an optimized variant referred to as YOLOv5p, which preserves core detection functionality while reducing computational and memory requirements. 
The model is quantized to 8-bit precision using GreenWaves' NNTool\footnote{\url{https://github.com/GreenWaves-Technologies/gap_sdk/tree/master/tools/nntool}}, which converts FP32 weights to a symmetric INT8 representation. 
\revA{This \cgls{PTQ} applies an affine transformation calibrated on representative data to determine dynamic ranges for each tensor. NNTool supports per-channel quantization of weights and per-tensor quantization of activations. To further optimize execution, the tool applies static graph optimization such as layer fusion and leverages an expression compiler to merge broadcast-friendly operations into efficient, low-overhead kernels.}
The model is then mapped onto the GAP9 \cgls{soc} using GreenWaves' AutoTiler, which performs operator tiling and code generation for hardware acceleration via the NE16 unit.
\revA{AutoTiler statically partitions operations and manages memory movement across the L1–L3 hierarchy, generating highly optimized C code with triple buffering.}
\begin{figure}[t]
    \centering
    \setlength{\abovecaptionskip}{-10pt}
    \includegraphics[width=1\columnwidth]{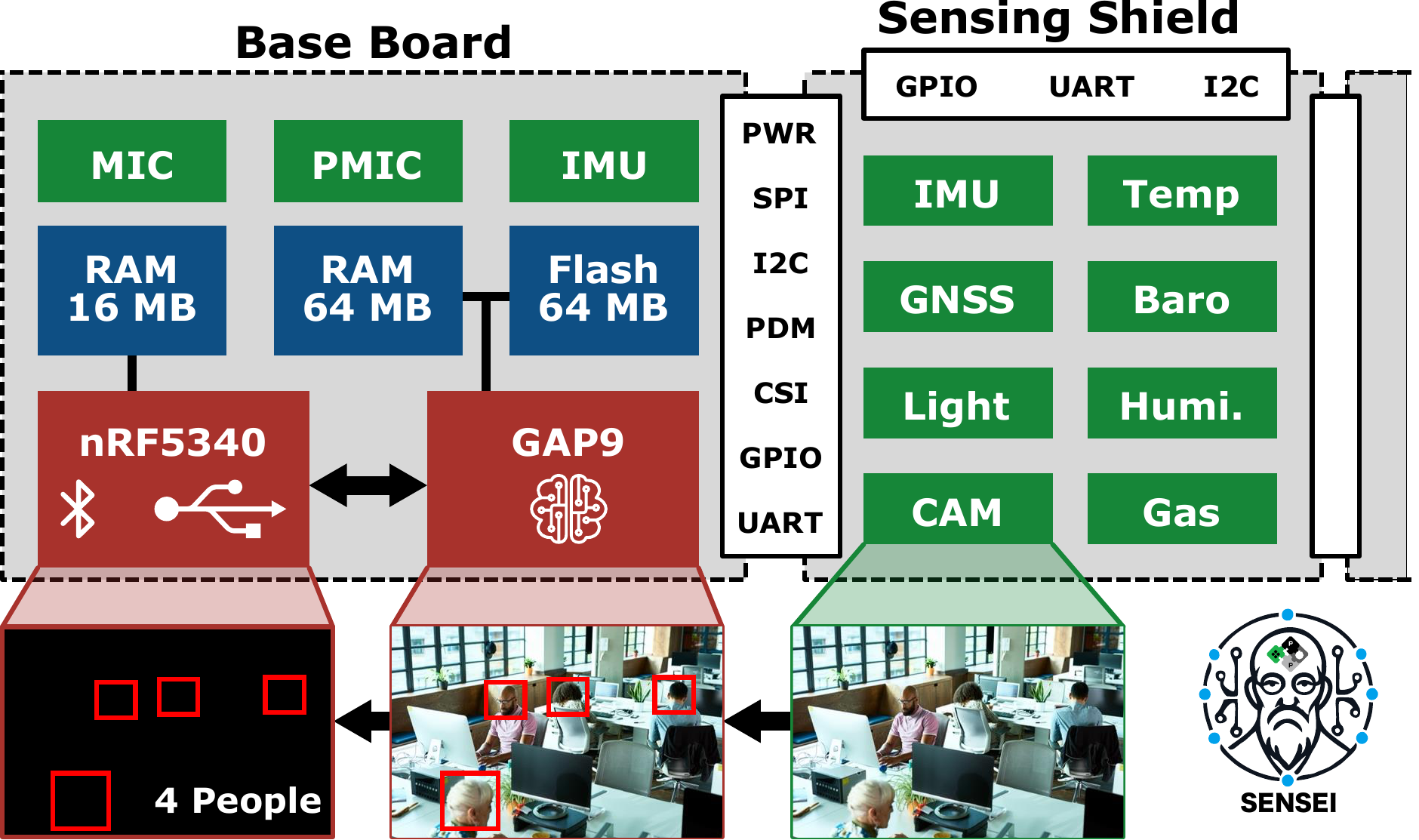}
    \caption{System setup for on-device occupancy detection. The camera is directly connected to GAP9, which performs debayering, image scaling, neural network inference, and post-processing. Final bounding boxes are extracted on GAP9 via \acrshort{nms} and transmitted over UART to the nRF, which also reads out environmental sensors.}
    \label{fig:implementation_occupancy}
\end{figure}

The on-device occupancy detection pipeline is shown in \Cref{fig:implementation_occupancy}.
An RGB image is captured by the camera and transferred directly to GAP9, where several pre-processing steps are performed, including debayering, auto white balancing, and image downscaling.
Following pre-processing, the quantized YOLOv5p model is executed using the hardware accelerator to perform inference on the processed image.
\revB{Post-processing is performed entirely on GAP9 and includes \cgls{nms} to keep only the highest-confidence bounding box for each detected head.}
The final output consists of the bounding box coordinates and corresponding confidence scores.
Occupancy is estimated by counting the remaining bounding boxes after suppression, which can be transmitted via UART to the nRF5340.

\subsection{Network Configuration} 

We train and evaluate multiple variants of the YOLOv5p architecture using RGB input images with different sizes from a public human head dataset~\cite{dataset-human-head_dataset}. 
The evaluation focuses on the trade-offs between detection accuracy and energy consumption per frame, as summarized in \Cref{tab:power_256}.
We measure the power consumption of GAP9, external memory, and the camera. 
We exclude the remaining sensors and the nRF from the measurement to isolate the energy required for occupancy detection alone.
All models share the same quantized backbone but differ in input resolution, ranging from \R{\SI[parse-numbers=false]{64\times64}{}} to \R{\SI[parse-numbers=false]{512\times512}{}}, which directly affects computational complexity and detection performance.
Each model is trained for \R{\SI{200}{epochs}} using stochastic gradient descent (SGD).
\input{tab/energy_breakdown}

\begin{figure}[t]
    \centering
    \subfloat[
        Time series of the power consumption during occupancy detection with \R{\SI[parse-numbers=false]{512\times512}{}} images on GAP9.
        \label{fig:timeseries}
    ]{
        \includegraphics[width=1\columnwidth]{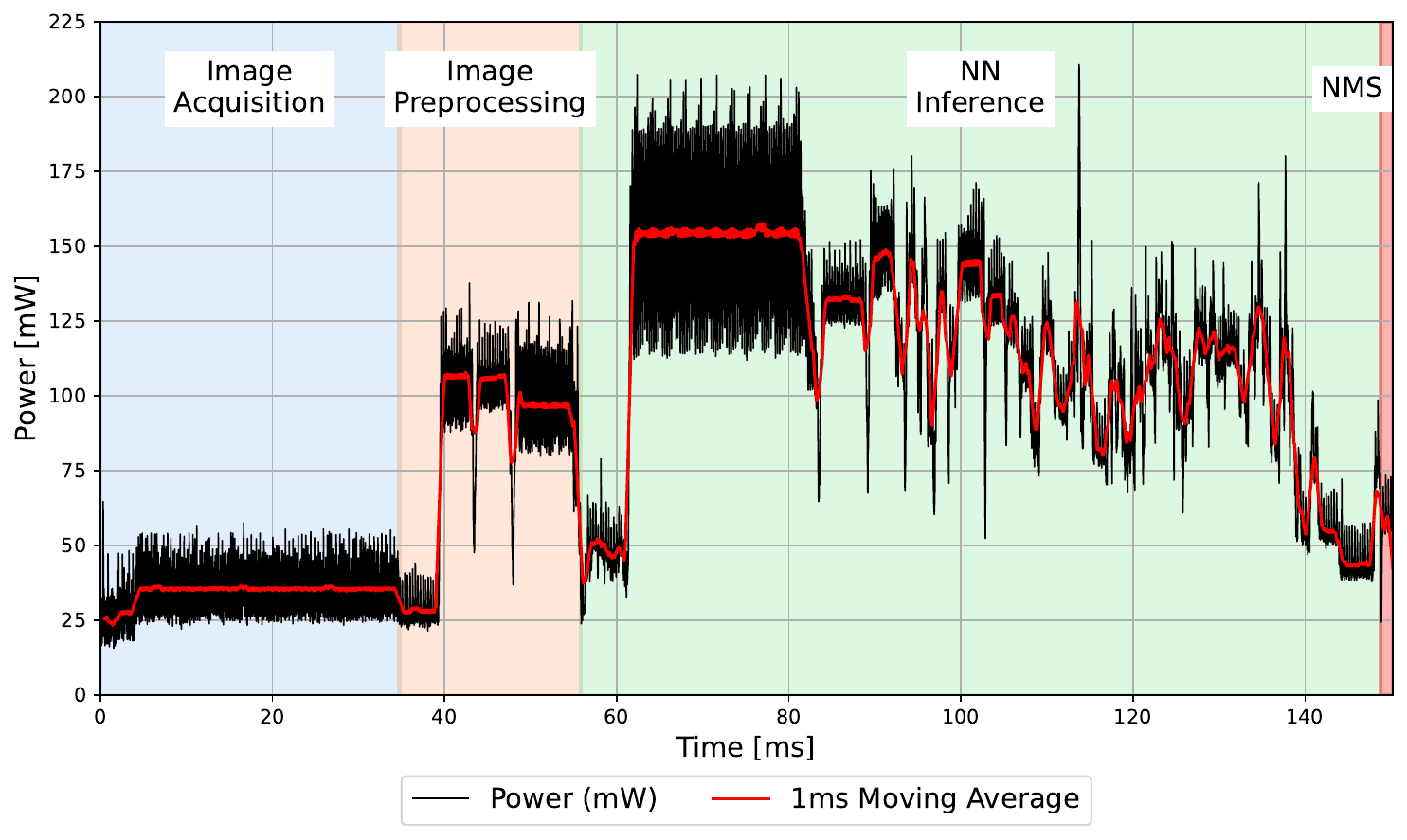}
    }
        
    \hfill
    \subfloat[
        Trade-off between input resolution, accuracy, energy per frame, and inference efficiency. Efficiency is calculated as the number of percentage points of \acrshort{mAP50}, normalized by the energy consumed per frame.
        \label{fig:tradeoff_plot}
    ]{
        \includegraphics[width=1\columnwidth]{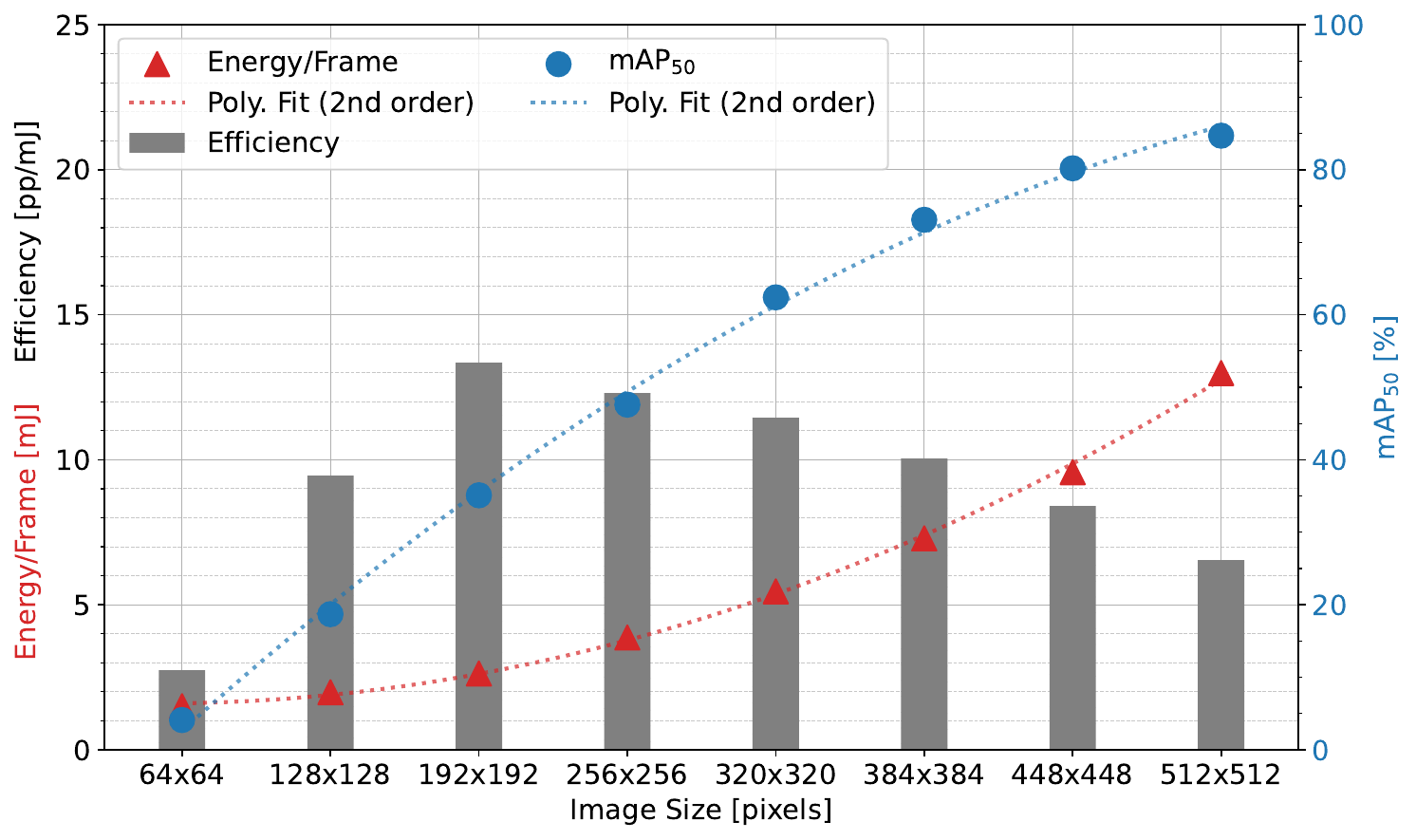}
    }
    \caption{Evaluation of runtime performance and energy efficiency across YOLOv5p configurations on GAP9.}
    \label{fig:energy_tradeoffs}
\end{figure}

\revB{The runtime characteristics of the \R{\SI[parse-numbers=false]{512\times512}{}} configuration are visualized in the power trace shown in \Cref{fig:timeseries}.}
To identify the most efficient implementation, we compute efficiency as accuracy (\acrshort{mAP50}) per energy consumed per frame. 
The efficiency depicted in \Cref{fig:tradeoff_plot} through the dashed-grey bars shows that the \R{\SI[parse-numbers=false]{192\times192}{}} configuration is most efficient with \R{\SI{13.6}{pp \per\milli\joule}}.

\subsection{Long-Term Operations} 

Considering the most effective configuration (\R{\SI[parse-numbers=false]{192\times192}{}}), we estimate the long-term deployment performance assuming one sample every \R{\SI{2}{\second}}, suitable for indoor occupancy monitoring.
Assuming maximum \cgls{BLE} throughput, a transmission power of \R{\SI{10}{\milli\watt}}, and a conservative sleep power of \R{\SI{1}{\milli\watt}}, we estimate a per-sample energy cost of \R{\SI{4.6}{\milli\joule}}. 
Using a \R{\SI{600}{\milli\amperehour}} battery at \R{\SI{3.7}{\volt}} results in over \R{40} days of uninterrupted operation. 

\revA{To contextualize these savings, we compare against a baseline that transmits raw \R{\SI[parse-numbers=false]{320\times240}{}} grayscale images over \cgls{BLE}, without any local processing or inference. 
Additionally, we do not account for the additional energy required to process the images off-device.
While this baseline configuration is inefficient in terms of transmission energy, it reflects a worst-case setup where no computational effort is placed on the device.
While compression techniques like JPEG would reduce the transmission load, they require non-negligible on-device computation and should therefore be considered a form of on-device processing.}
\revB{This baseline setup results in} \R{\SI{7.86}{\milli\joule}} per sample, reducing the operational lifetime to only \R{24} days.
This \R{\SI{42}{\percent}} reduction in energy consumption is primarily achieved by eliminating high-bandwidth wireless transmissions and leveraging on-device computation. 
By performing acquisition, inference, and filtering directly on GAP9, the average power draw drops from \R{\SI{3.93}{\milli\watt}} (excluding the power datacenter power) to \R{\SI{2.29}{\milli\watt}}, highlighting the benefits of edge intelligence for resource-constrained \cgls{IoT} applications.

%% file: tab/energy_breakdown.tex
\begin{table*}[tb]
    \centering
    \setlength{\abovecaptionskip}{-8pt}
    \caption{Accuracy, model parameter, power breakdown, and energy per frame for indoor occupancy detection using YOLOv5p.}
    \begin{threeparttable}
    \label{tab:power_256}
    \setlength{\tabcolsep}{11pt}
    \begin{tabular}{@{}c*2S[table-format=3.1]S[table-format=3]S[table-format=3]*6S[table-format=2.1]@{}}
        \toprule
        \textbf{Image Size} 
        & \textbf{mAP\textsubscript{50}\tnote{*}} 
        & \textbf{mAP\textsubscript{50-95}\tnote{\textdagger}} 
        & \textbf{Op} & \textbf{Params} 
        & \textbf{GAP9} & \textbf{Memory} & \textbf{Camera} & \textbf{Total} 
        & \textbf{FPS} & \textbf{Energy/Frame} \\
        
        [pixel] 
        & [\si{\percent}] 
        & [\si{\percent}] 
        & [\si{M}] & [\si{K}] 
        & [\si{\milli\watt}] & [\si{\milli\watt}] & [\si{\milli\watt}] & [\si{\milli\watt}] 
        & [\si{1\per\second}] & [\si{\milli\joule}] \\
        
        \midrule
        64$\times$64
        & 4.1 
        & 1.1
        & 5 & 309
        & 23.5 & 0.8 & 13.5  & 37.8 
        & 25.2 & 1.50 \\
        128$\times$128
        & 18.7 
        & 6.7
        & 19 & 312 
        & 31.2 & 0.8 & 11.8 & 43.8
        & 22.2 & 1.98 \\
        \bfseries 192$\times$192
        & \bfseries 35.1 
        & \bfseries 13.4 
        & \bfseries 42 & \bfseries 317 
        & \bfseries 40.1 & \bfseries 0.8 & \bfseries 10.4 & \bfseries 51.3
        & \bfseries 19.5 & \bfseries 2.63 \\
        256$\times$256
        & 47.6 
        & 19.5 
        & 74 & 324 
        & 49.4 & 1.1 & 8.1 & 58.5
        & 15.1 & 3.87 \\
        320$\times$320
        & 62.4 
        & 25.5 
        & 116 & 333 
        & 58.7 & 1.3 & 6.5 & 66.5
        & 12.2 & 5.46 \\
        384$\times$384
        & 73.1 
        & 30.4 
        & 167 & 345
        & 64.2 & 1.6 & 5.2 & 71.0
        & 9.7 & 7.29 \\
        448$\times$448
        & 80.2
        & 34.9
        & 227 & 358
        & 65.6 & 5.6 & 4.2 & 75.4
        & 7.9 & 9.57 \\
        \bfseries 512$\times$512
        & \bfseries 84.7
        & \bfseries 37.7
        & \bfseries 297 & \bfseries 373
        & \bfseries 73.9 & \bfseries 8.9 & \bfseries 3.6 & \bfseries 86.4
        & \bfseries 6.7 & \bfseries 12.99 \\
        \bottomrule
    \end{tabular}
    \begin{tablenotes}
    \item [*] \Acrlong{mAP50}
    \item [\textdagger] \Acrlong{mAP5095}
    \end{tablenotes}
    \end{threeparttable}
    \vspace{-2em}
\end{table*}

%% file: src/50_Results.tex
\section{Smart Environmental Monitoring}
\label{sec:result}


Our platform is designed to support long-term, self-sufficient, and privacy-aware \cgls{EM} by combining multi-modal sensing with embedded edge processing. 
It enables high-resolution measurements of physical parameters, such as \ce{CO2}, \cglspl{voc}, and temperature, while integrating occupancy data derived from on-device inference.
To demonstrate these capabilities, we assess the behavior of the sensor array and showcase the system’s full-stack functionality by combining occupancy detection (described in \Cref{sec:occupancy_detection}) with air quality monitoring.

\subsection{Sensor Characterization} \label{sec:result:sensor}

\begin{figure}[t]
    \centering
    \subfloat[
        Indoor office test showing increased \acrshortpl{voc} during occupancy, followed by a decrease upon activation of ventilation.
        \label{fig:test_indoors}
    ]{
        \includegraphics[width=1\columnwidth]{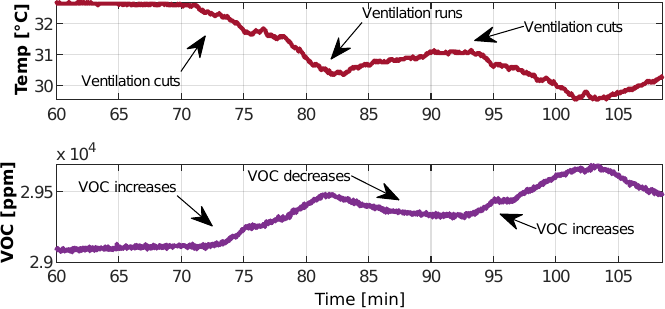}
    }
    \hfill
    \subfloat[
        Outdoor measurements from morning to midday, capturing trends in temperature, UV radiation, and barometric pressure.
        \label{fig:test_outdoors}
    ]{
        \includegraphics[width=1\columnwidth]{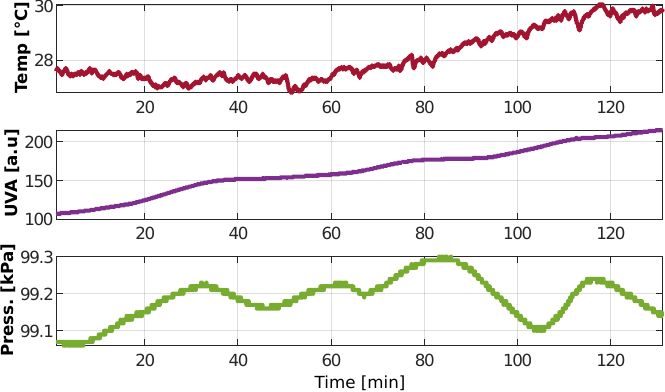}
    }
    \caption{Environmental sensor readings in two test scenarios: (a) indoor office with occupancy-driven variations, and (b) outdoor exposure capturing daily changes.}
    \label{fig:sensor_tests}
\end{figure}

We examine the behavior of the sensor array in two representative deployment scenarios: a shared indoor office and an outdoor setting. 
The indoor test captures air quality indicators influenced by human presence, including temperature and \cglspl{voc}, while the outdoor test records ambient light, UV radiation, and barometric pressure variations over the course of a morning.
\Cref{fig:test_indoors} illustrates sensor responses in a shared office environment occupied by three people. 
During occupancy, levels of \cglspl{voc} increase over time, followed by a notable decrease when ventilation is activated. 
\Cref{fig:test_outdoors} shows outdoor data collected from early morning until midday, with expected rises in light intensity, UV exposure, and temperature, confirming the sensors’ responsiveness to environmental changes.

Across both scenarios, we measure the total energy of \R{\SI{490}{\milli\joule}} required to read out all sensors from the nRF.
The \textit{SCD41}, \textit{BME680}, and \textit{SGP41} are the most power-intensive components, drawing up to \R{\SI{160}{\milli\watt}} due to internal heating elements. 
In contrast, the remaining sensors operate below \R{\SI{30}{\milli\watt}}, making them suitable for frequent sampling in long-term deployments.
Sensors can be sampled less frequently based on contextual triggers such as occupancy or time of day to improve overall energy efficiency.

\subsection{End-To-End Application}



In the full deployment scenario, we select the \R{\SI[parse-numbers=false]{512\times512}{}} network occupancy detection network for a fair comparison with related work~\cite{yen2023adaptive}.
The occupancy detection runs at a \R{\SI{2}{\second}} interval, while the environmental sensors are sampled once every \R{\SI{60}{\second}}.
Although the actual sampling frequency of the sensors can be reduced during periods of inactivity, we conservatively assume that the room remains occupied and use a sampling interval of \R{\SI{60}{\second}} for the following estimation. 
We assume a system-wide sleep power of \R{\SI{1}{\milli\watt}} between sampling events. 
Under these conditions, the total energy required for one complete cycle, including image acquisition, inference, and full sensor readout, is \R{\SI{929}{\milli\joule}}, resulting in an average power consumption of \R{\SI{15.5}{\milli\watt}}.
This corresponds to a runtime of almost \R{\SI{6}{days}} on a standard \R{\SI{600}{\milli\amperehour}}, \SI{3.7}{\volt} battery. 
In addition to energy-efficient processing, the system provides several critical advantages: it preserves privacy by processing data locally, eliminates dependency on cloud infrastructure, and reduces system cost.

%% file: src/70_Discussion.tex
\begin{table*}[t]
\centering
\setlength{\abovecaptionskip}{-0pt}
\caption{Comparison of SoA environmental \cgls{IoT} monitoring systems with embedded processing capabilities}
\renewcommand{\arraystretch}{1.1}
\begin{threeparttable}
    \begin{tabularx}{\textwidth}{@{}lclXr@{}}
\toprule
    \textbf{Author} 
    & \textbf{Sensors} 
    & \textbf{Embedded Processor} 
    & \textbf{Application} 
    & \textbf{Power} \\
\midrule
Moursi et al. \cite{moursi2021iot}     
    & \SIrange{3}{5}{}        
    & Yes (Raspberry Pi 4) 
    & ML Inference (Centralized Edge) & $\sim$\SI{4.5}{\watt}\tnote{*} \\
Albanese et al. \cite{albanese2021automated} 
    & 1 
    & Yes (Raspberry Pi 3 + NCS2) 
    & DNN-Based Pest Detection & $\sim$\SI{4.5}{\watt}\tnote{*} \\
Shadrin et al. \cite{Shadrin2020EmbeddedAI}   
    & 1       
    & Yes (Raspberry Pi 3B + NCS2)      
    & \cgls{lstm} Plant Modeling   & $\sim$\SI{6.0}{\watt}\tnote{*} \\
Yen et al. \cite{yen2023adaptive} 
    & 1 
    & Yes (Raspberry Pi 4) 
    & CNN-Based Occupancy every \SI{0.5}{\second} 
    & $\sim$\SI{6.0}{\watt}~ \\
\textbf{This work}     
    & \textbf{11} 
    & \textbf{Yes (nRF5340 + GAP9)} 
    & \textbf{CNN-Based Occupancy every \SI{2}{\second} + Air Quality Monitoring every \SI{60}{\second}} 
    & \textbf{\SI{15.5}{\milli\watt}}~ \\
\bottomrule
\end{tabularx}
\begin{tablenotes}
\footnotesize
\item[*] Power values estimated from typical operating conditions: Raspberry Pi 3B ($\sim$\SI{3}{\watt}), Raspberry Pi 4 ($\sim$\SI{4.5}{\watt}), Intel NCS2 ($\sim$\SI{1.5}{\watt}).
\end{tablenotes}
\label{tab:soa_comparison}
\end{threeparttable}
\vspace{-2em}
\end{table*}

\section{Discussion}
\label{sec:discussion}


\Cref{tab:soa_comparison} compares our system with representative state-of-the-art environmental monitoring platforms that incorporate embedded \cgls{ai} capabilities.
While prior work has demonstrated the feasibility of on-device inference for \cgls{EM} tasks, these systems mostly rely on general-purpose single-board computers such as Raspberry~Pi 3 or 4, often paired with external \cgls{ai} accelerators like the Intel NCS2. 
Although powerful, these platforms consume several watts of power and are unsuitable for long-term battery operation.

In contrast, our system combines multiple distinguishing features. 
First, it supports efficient on-device \cgls{ai} inference through the GAP9 \cgls{soc}.
It offers highly flexible and hardware-accelerated execution of quantized neural networks at ultra-low power with fine-grained control over memory and execution schedules. 
Second, it integrates 11 environmental sensors into a compact sensing platform, including gas, motion, and light sensors. 
Thirdly, the \cgls{mcu}-based system achieves smart autonomous \cgls{iaq}, operating orders of magnitude below comparable Raspberry Pi-based solutions.
Compared to~\cite{yen2023adaptive}, we achieve more than \R{$380\times$} lower power consumption while integrating a broader range of sensors within a smaller, battery-powered platform.
Finally, it preserves privacy by keeping all image data local to the device, avoiding transmitting sensitive information. 
It enables self-reliance by operating without external compute or storage infrastructure, reducing latency and system complexity. 
This lowers the overall system cost and enhances robustness and safety in critical monitoring applications, such as in building automation, healthcare, or resource-constrained environments.

%% file: src/80_Conclusion.tex
\section{Conclusion}
\label{sec:conclusion}
We introduced a compact, energy-efficient \cgls{IoT} platform for multi-modal environmental monitoring that combines embedded machine learning with a dual-\cgls{soc} architecture based on GAP9 and nRF5340. 
Our design integrates four key features: support for on-device AI inference, flexible hardware acceleration for quantized neural networks, comprehensive multi-modal sensing with 11 heterogeneous sensors, and an ultra-low-power \cgls{mcu}-based implementation. 
Unlike prior solutions that rely on power-hungry single-board computers, our platform operates fully autonomously at just \R{\SI{15.5}{\milli\watt}}, enabling real-time occupancy detection and environmental sensing with a runtime of over \R{\SI{143}{\hour}} on a \R{\SI{600}{\milli\amperehour}} battery. 
In addition to its energy efficiency, the system ensures privacy by keeping all data local, eliminates the need for cloud infrastructure, and provides a scalable and robust foundation for future autonomous monitoring and predictive control applications.
